\documentclass[sigplan,10pt]{acmart}
\startPage{1}

\setcopyright{none}
\usepackage{booktabs}   %% For formal tables:
\usepackage{subcaption} %% For complex figures with subfigures/subcaptions
\usepackage{algorithm}
\usepackage{algpseudocode}
\usepackage{listings}  % for your lstlist
\usepackage{natbib}
\usepackage{graphicx}

\usepackage{microtype}        % Better character spacing
\usepackage[english]{babel}   % Better hyphenation
\providecommand{\tightlist}{%
  \setlength{\itemsep}{0pt}\setlength{\parskip}{0pt}}



\begin{document}

%% Title information
\title[CMP Paper]{No Cords Attached: Coordination-Free Concurrent Lock-Free Queues}         %% [Short Title] is optional;
                                        %% when present, will be used in
                                        %% header instead of Full Title.
%\titlenote{For simple use only}             %% \titlenote is optional;
                                        %% can be repeated if necessary;
                                        %% contents suppressed with 'anonymous'
%\subtitle{Check the format}                     %% \subtitle is optional
%\subtitlenote{Everything should just work}       %% \subtitlenote is optional;
                                        %% can be repeated if necessary;
                                        %% contents suppressed with 'anonymous'

%% Author information
%% Contents and number of authors suppressed with 'anonymous'.
%% Each author should be introduced by \author, followed by
%% \authornote (optional), \orcid (optional), \affiliation, and
%% \email.
%% An author may have multiple affiliations and/or emails; repeat the
%% appropriate command.
%% Many elements are not rendered, but should be provided for metadata
%% extraction tools.

%% Author with single affiliation.
\author{Yusuf Motiwala}
% \authornote{with author1 note}          %% \authornote is optional;
%                                         %% can be repeated if necessary
\orcid{0009-0004-8038-1110}             %% \orcid is optional
\affiliation{
%   \position{Founder}
%   \department{Department1}              %% \department is recommended
\institution{mesibo and PatANN}            %% \institution is required
%   \streetaddress{Street1 Address1}
%   \city{City1}
%   \state{State1}
%   \postcode{Post-Code1}
\country{}                    %% \country is recommended
}
\email{yusuf@mesibo.com}          %% \email is recommended

% %% Author with two affiliations and emails.
% \author{First2 Last2}
% \authornote{with author2 note}          %% \authornote is optional;
%                                         %% can be repeated if necessary
% \orcid{nnnn-nnnn-nnnn-nnnn}             %% \orcid is optional
% \affiliation{
%   \position{Position2a}
%   \department{Department2a}             %% \department is recommended
%   \institution{Institution2a}           %% \institution is required
%   \streetaddress{Street2a Address2a}
%   \city{City2a}
%   \state{State2a}
%   \postcode{Post-Code2a}
%   \country{Country2a}                   %% \country is recommended
% }
% \email{first2.last2@inst2a.com}         %% \email is recommended
% \affiliation{
%   \position{Position2b}
%   \department{Department2b}             %% \department is recommended
%   \institution{Institution2b}           %% \institution is required
%   \streetaddress{Street3b Address2b}
%   \city{City2b}
%   \state{State2b}
%   \postcode{Post-Code2b}
%   \country{Country2b}                   %% \country is recommended
% }
% \email{first2.last2@inst2b.org}         %% \email is recommended

%% Remove these two lines for the camera-ready version
\fancyhead{}  % clears the header fields that trigger HotCRP format checker
\renewcommand\footnotetextcopyrightpermission[1]{} % removes the ACM copyright footnote

%% Abstract
%% Note: \begin{abstract}...\end{abstract} environment must come
%% before \maketitle command
\begin{abstract}
%%Abstract here \ldots
The queue is conceptually one of the simplest data structures—a basic FIFO container. However, ensuring correctness in the presence of concurrency makes existing lock-free implementations significantly more complex than their original form. Coordination mechanisms introduced to prevent hazards such as ABA, use-after-free, and unsafe reclamation often dominate the design, overshadowing the queue itself. Many schemes compromise strict FIFO ordering, unbounded capacity, or lock-free progress to mask coordination overheads. Yet the true source of complexity lies in the pursuit of infinite protection against reclamation hazards—theoretically sound but impractical and costly. This pursuit not only drives unnecessary complexity but also creates a protection paradox where excessive protection reduces system resilience rather than improving it. While such costs may be tolerable in conventional workloads, the AI era has shifted the paradigm: training and inference pipelines involve hundreds to thousands of concurrent threads per node, and at this scale, protection and coordination overheads dominate, often far heavier than the basic queue operations themselves.
	
This paper introduces Cyclic Memory Protection (CMP), a coordination-free queue that preserves strict FIFO semantics, unbounded capacity, and lock-free progress while restoring simplicity. CMP reclaims the strict FIFO that other approaches sacrificed through bounded protection windows that provide practical reclamation guarantees. We prove strict FIFO and safety via linearizability and bounded reclamation analysis, and show experimentally that CMP outperforms state-of-the-art lock-free queues by up to 1.72-4× under high contention while maintaining scalability to hundreds of threads. Our work demonstrates that highly concurrent queues can return to their fundamental simplicity without weakening queue semantics.
\end{abstract}

%% \maketitle
%% Note: \maketitle command must come after title commands, author
%% commands, abstract environment, Computing Classification System
%% environment and commands, and keywords command.
\maketitle

%% \section{Introduction}
%% The paper contents here \ldots

% Auto-generated input sequence
\hypertarget{introduction}{%
\section{Introduction}\label{introduction}}

Modern workloads often demand three properties from concurrent queues
simultaneously: (1) unbounded capacity, to absorb bursty or
unpredictable workloads without artificial limits; (2) strict FIFO
ordering, so elements are dequeued in the exact order they were
enqueued, regardless of which thread performs the operation; and (3)
scalable performance, so per-operation cost remains constant as thread
count grows, with lock-free progress ensuring one operation never stalls
another.

Despite decades of research, existing non-blocking (lock-free or
wait-free) queue algorithms still trade among throughput, ordering, and
memory safety. High-throughput designs often relax FIFO ordering or
impose fixed capacity. A central obstacle is the ABA problem: when a
thread reads a value A, another thread changes it to B and then back to
A, and the first thread later performs a CAS that succeeds under the
false assumption that nothing has changed. This hazard can cause stale
pointers to be mistaken for valid ones, leading to use-after-free (UAF)
errors or corrupted states in both linked and array-based designs.
Preventing ABA without prohibitive synchronization remains a major
challenge, especially in unbounded queues where nodes are frequently
allocated and freed.

We introduce Cyclic Memory Protection (CMP), a fully lock-free queue
where enqueue, dequeue, and reclamation proceed concurrently without
blocking or coordination. CMP provides formal safety guarantees,
tolerates stalled or failed threads without delaying progress, and makes
the following contributions:

\begin{itemize}
\tightlist
\item
  \textbf{Lock-free progress with unbounded capacity.} Supports bursty
  and unpredictable workloads without artificial limits.
\item
  \textbf{Strict FIFO ordering.} Preserves the exact global enqueue
  order across all threads, regardless of which thread performed the
  enqueue.
\item
  \textbf{Scalable performance.} Achieves constant per-operation cost
  under contention.
\item
  \textbf{Memory safety.} Provides formal guarantees against ABA and UAF
  hazards.
\item
  \textbf{Bounded reclamation.} Provides guaranteed reclamation within
  finite cycles, unlike hazard pointers or epoch-based schemes that may
  indefinitely retain nodes.
\item
  \textbf{Fault tolerance.} Ensures progress and reclamation despite
  stalled or failed threads, without global pauses or manual cleanup.
\end{itemize}

To our knowledge, no existing production-ready queue achieves all of
these properties together with comparable performance and scalability.

\hypertarget{background-and-prior-work}{%
\section{Background and Prior Work}\label{background-and-prior-work}}

\hypertarget{history-of-lock-free-queues}{%
\subsection{History of Lock-Free
Queues}\label{history-of-lock-free-queues}}

The Michael and Scott linked-list queue \citep{michael1996simple}
introduced a simple, lock-free, multi-producer/multi-consumer design
that remains a template for many modern concurrent queues. While it
solved the basic synchronization problem, it left a critical correctness
gap: the ABA problem.

For nearly a decade, lock-free queues saw limited adoption due to unsafe
memory management. Deployment required either accepting memory leaks,
relying on garbage-collected languages, or adopting ad hoc, often unsafe
schemes. In 2004, Maged Michael, one of the original queue's authors,
introduced hazard pointers \citep{michael2004hazard}, the first widely
adopted reclamation scheme for lock-free data structures. This was
followed by a series of other reclamation techniques, each attempting to
address the limitations of its predecessors.

\hypertarget{memory-reclamation-and-coordination-complexity}{%
\subsection{Memory Reclamation and Coordination
Complexity}\label{memory-reclamation-and-coordination-complexity}}

Michael's hazard pointers prevents ABA problems by requiring threads to
publish the pointers they are currently accessing in shared hazard
pointer slots. Before reclaiming any retired object, a thread must scan
all hazard pointer slots across all threads to ensure no other thread
holds a reference. This requires \(O(P \times K)\) comparisons per
reclamation pass, where \texttt{P} is the number of threads and
\texttt{K} is the number of hazard slots per thread. This scan
introduces coordination costs that scale linearly with \texttt{P}, and
frequent updates to hazard slots can cause cache-line contention and
memory-barrier overheads that can be prohibitive under modern multicore
systems with hundreds or even thousands of threads.

These coordination bottlenecks and overheads motivated alternative
approaches including Quiescent-State-Based Reclamation (QSBR)
\citep{mckenney1998read}, reference counting schemes
\citep{valois1995lock}, tagged pointer approaches \citep{ibm1983system},
Pass The Buck methods \citep{herlihy2005ptb}, Epoch-Based Reclamation
(EBR), distributed epoch-based techniques like DEBRA
\citep{brown2015reclaiming} and more. However, each of these approaches
either introduces different coordination overheads, suffers from
unpredictable reclamation delays, or faces fundamental scalability
limitations as thread counts increase.

Epoch-Based Reclamation (EBR) batches retired nodes into epochs and
reclaims only after all threads advance past the target epoch. This
amortizes coordination to \texttt{O(P)} but makes reclamation depend on
the slowest (or crashed) thread, causing unbounded retention; if
allocation or reclamation lies on the hot path, this can impede
system-level lock-free progress under stalls. Distributed variants such
as DEBRA reduce overhead with per-thread retire lists and lighter
synchronization, but reclamation still requires every thread to pass a
quiescent state, so a stalled participant can delay frees
\citep{brown2015reclaiming}.

QSBR/RCU rely on application quiescent points; they work well when
threads cooperate, but guarantees weaken outside that model
\citep{mckenney1998read}. Reference counting prevents use-after-free but
not ABA under reuse and also adds per-access atomic overhead
\citep{valois1995lock}. Tagged/sequence pointers help detect stale CAS
values (ABA) but do not prevent premature reuse; larger tags reduce
wraparound risk at the cost of wider atomics and still require a
reclamation mechanism \citep{michael2004hazard, brown2015reclaiming}.

Beyond the raw costs, threads must constantly validate that their
protection remains valid, handle failures gracefully, and maintain
complex retry state machines. The overhead of these coordination-heavy
protocols exceeds that of the core queue operations themselves, making
memory safety more expensive than the data structure it protects.
Schemes like reference counting, although based on a different design,
incurs high per-access atomic overhead and is typically slower than
coordinated schemes.

A comprehensive survey by Hart et al. \citep{hart2007performance}
demonstrates that existing reclamation schemes struggle to achieve both
safety and performance at scale, motivating the search for truly
coordination-free alternatives.

\hypertarget{the-scalability-paradox-in-memory-reclamation}{%
\subsection{The Scalability Paradox in Memory
Reclamation}\label{the-scalability-paradox-in-memory-reclamation}}

While these techniques made important progress toward addressing ABA and
UAF hazards in concurrent data structures, they introduced a different
class of problems: the coordination itself became a source of
complexity, fragility, and performance bottlenecks. What begins as a
mechanism to ensure safety ends up creating new correctness and
scalability challenges --- the scalability paradox.

\hypertarget{fragility-under-thread-failure}{%
\subsubsection{Fragility under Thread
Failure}\label{fragility-under-thread-failure}}

Despite these layers of coordination, the schemes remain fragile under
thread failure. All these protocols rest on a common assumption that
every thread will eventually participate, but in practice, threads may
stall or crash unexpectedly. A stalled hazard pointer can prevent
reclamation permanently, a crashed participant can halt epoch
advancement, and missing quiescent state reports can extend grace
periods indefinitely. Even distributed schemes like DEBRA remain
vulnerable to group blocking when a member fails. Attempts to work
around this through watchdogs, external monitors, or heartbeat
mechanisms \citep{dice2016recoverable} introduce additional
infrastructure requirements, moving parts, and heuristic-based tuning.
These measures add complexity and fragility, compounding the paradox
rather than resolving it.

\hypertarget{real-world-queue-designs-and-trade-offs}{%
\subsubsection{Real-World Queue Designs and
Trade-offs}\label{real-world-queue-designs-and-trade-offs}}

To work around these limitations and the performance penalty, modern
high-throughput queue designs often relax one or more fundamental
properties. The result is a spectrum of trade-offs, where scalability is
gained at the expense of FIFO ordering, unbounded capacity, or
lock-freedom.

Moodycamel's ConcurrentQueue achieves excellent performance by using
per-producer segmented subqueues. However, this comes at the cost of
strict FIFO: ordering is preserved only within each producer, while
interleaving between producers is permitted. Vyukov's queue delivers
near-O(1) operations with strict per-slot FIFO but requires capacity to
be fixed at initialization, sacrificing unboundedness. General-purpose
concurrency frameworks such as Intel TBB and Meta's Folly retain both
FIFO and unbounded capacity by introducing fine-grained or hybrid locks,
but giving up lock-freedom and incurring blocking overhead under
contention.

Even with these trade-offs, these coordination-heavy reclamation
protocols become prohibitively expensive in modern many-core
environments requiring hundreds or thousands of concurrent threads.

The root cause is not the trade-off itself but the chosen protection
model. Existing schemes attempt to guarantee indefinite protection
against ABA and UAF, which is not only unrealistic in production but
also introduces complexity and fragility. This tension gives rise to
what is known as \textbf{the protection paradox}.

\hypertarget{the-protection-paradox}{%
\subsubsection{The Protection Paradox}\label{the-protection-paradox}}

To remain safe in the face of stalls, existing schemes adopt
conservative policies: nodes are not reclaimed until all threads
cooperate again. While this guarantees safety in theory, it rests on two
flawed assumptions.

First, they assume stalled threads will eventually recover. In
production systems, this assumption rarely holds. The consequence is
unbounded and unpredictable reclamation delays, leading to uncontrolled
memory retention and no automatic recovery from failed threads.

Second, they impose a hot-path tax on every operation to guard against
failures that occur less than 0.1\% of the time in production
deployments. This results in slower performance 99.9\% of the time, yet
those rare failures still stall reclamation indefinitely. A crashing or
stalled thread is a separate bug to fix---not a justification for taxing
the fast path 99.9\% of the time. Worse, ``infinite'' protection
inevitably leads to ``infinite'' leak duration.

This is the classic protection paradox: adding more protection against
rare failures makes the system less resilient and reduces practical
safety and scalability. It also introduces overhead greater than that of
the queue itself. Cyclic Memory Protection (CMP) takes the opposite
stance: it replaces indefinite protection with a bounded temporal safety
window, eliminating inter-thread coordination while tolerating stalled
or failed threads. This keeps the common case lean, bounds the uncommon
case, and provides predictable memory use and stable latency for
unbounded, strictly FIFO, lock-free queues.

\hypertarget{the-cmp-algorithm}{%
\section{The CMP Algorithm}\label{the-cmp-algorithm}}

Cyclic Memory Protection (CMP) is a lock-free MPMC queue that achieves
strict FIFO ordering and unbounded capacity with bounded reclamation
requiring no inter-thread coordination. CMP eliminates use-after-free
hazards and uses a configurable protection window to bound reclamation
and prevent ABA within the window. Building on the Michael \& Scott
linking discipline, CMP removes stale helping and enables fully
concurrent reclamation without per-thread announcements (hazard
pointers, epochs, etc.). A scan cursor and a single claim CAS make
enqueue and dequeue O(1) in the common case.

\hypertarget{dual-protection-mechanism-1}{%
\subsection[Dual Protection Mechanism ]{\texorpdfstring{Dual Protection
Mechanism \footnote{\textbf{Memory Ordering Convention:} \texttt{LOAD}
  uses acquire ordering for reads requiring visibility of prior writes;
  \texttt{STORE} uses relaxed ordering except for publishing operations
  which use release; \texttt{FETCH\_MIN} and \texttt{INCREMENT} use
  sequential consistency or relaxed; CAS operations use acquire-release
  semantics. Non-atomic operations are safe due to immutability (e.g.,
  \texttt{current.cycle}) or single-writer guarantees (e.g., recovery
  state updates).}}{Dual Protection Mechanism }}\label{dual-protection-mechanism-1}}

The key idea is that protection is established before any thread
dereferences a node. A sliding protection window defines which nodes are
temporally safe; within this window, nodes cannot be reclaimed, so
threads dereference without handshakes or coordination. After accessing
a node, a thread unilaterally publishes its observation point (the
node's cycle), sliding the window forward for future nodes. Safety
against ABA and use-after-free (UAF) follows from mathematical
invariants, not consensus.

This preemptive model contrasts with traditional reactive approaches
(hazard pointers, epochs) that attempt protection after dereferencing,
often leading to additional complex race conditions and retries. CMP
eliminates this by establishing temporal safety boundaries first,
accessing once, and moving on, resulting in scalable performance with
fewer atomic operations and lower contention.

CMP ensures safety through two independent but complementary mechanisms:

\textbf{1. State-based protection.} Each node follows a two-state
lifecycle: \texttt{AVAILABLE\ →\ CLAIMED}. \texttt{AVAILABLE} nodes are
protected from reclamation and eligible for dequeue operations.
\texttt{CLAIMED} nodes become reclamation candidates once they satisfy
the cycle-based protection condition.

\textbf{2. Cycle-based sliding window protection.} Each enqueued node is
assigned an immutable monotonically increasing \texttt{cycle} at
creation, establishing a temporal identity that persists throughout its
lifetime. Dequeues publish the cycle of the node they claim as
\texttt{deque\_cycle}, which establishes an active sliding protection
window, \texttt{P\ =\ {[}deque\_cycle\ -\ W,\ deque\_cycle{]}}

where \texttt{W} is the protection window size, tuned to balances memory
usage with tolerance to thread delays,
\texttt{W\ =\ max\ (MIN\_WINDOW,\ OPS} \(\times\) \texttt{R)} where
\texttt{OPS} is expected dequeue rate (ops/s) and \texttt{R} is
resilience in seconds (maximum acceptable thread delay). The window
should accommodate the worst-case delay between dequeues under normal
and stressed conditions. Larger windows increases resilience to
scheduling delays and preemption but increase memory use, which is
bounded by \texttt{window\_size} \(\times\) \texttt{node\_size}
regardless of total queue capacity. \texttt{W} is configured per queue
instance at initialization, allowing different queues in the same
deployment to use different window sizes based on their workload
characteristics.

The key insight is that dequeue operations collectively maintain a
protection boundary without coordination, guaranteeing safe access to
future nodes and even to \texttt{CLAIMED} nodes that stalled threads may
still observe. When a new dequeue operation begins, protection from
previous dequeues already exists, enabling immediate safe access without
per-operation coordination. This preemptive approach eliminates the
coordination overhead, complex retry logic, and timing dependencies
common in traditional memory reclamation schemes, making dequeue latency
predictable and scalable regardless of thread count or timing anomalies.

While state-based protection is common in lock-free algorithms, the
sliding protection window approach eliminates the need for inter-thread
coordination entirely. Each node receives a cycle timestamp at enqueue,
and garbage collection maintains a protection window \texttt{P} based on
the highest cycle accessed by any dequeue operation. Nodes can only be
reclaimed when they are both \texttt{CLAIMED} and outside this sliding
window. This cycle-based protection provides bounded memory usage with
configurable parameters, automatic advancement without coordination, and
resilience to thread failures - unlike traditional approaches that
require hazard pointer scanning, epoch coordination, or other complex
synchronization protocols.

These two mechanisms together provide safety that is both mathematically
robust and operationally simple. State protection prevents freeing any
node that may still be accessed, while the cycle window adds a temporal
guard that covers stalls and failures. In combination, they eliminate
UAF and ABA by construction, without auxiliary validation protocols.
This redundancy does not add overhead: the mechanism relies only on the
atomic operations already required for correctness.

The dual protection also improves performance. Each operation is
self-contained and coordination-free, which improves cache locality and
delivers linear scalability across cores. It additionally provides
automatic recovery: coordinated schemes can let a stalled thread block
reclamation and trigger unbounded memory growth, whereas the cycle
window naturally expires old protection so a stalled thread cannot hold
the system hostage. This behavior is crucial in production, where long
preemptions and failures are inevitable.

\hypertarget{data-structure-and-rationale}{%
\subsection{Data Structure and
Rationale}\label{data-structure-and-rationale}}

The data structure design implements a lock-free FIFO queue with state-
and cycle-based protection for memory safety. It comprises two
components: a lock-free queue structure that handles FIFO linking and
node state transitions, and a cycle management structure that enforces
temporal ordering and controls reclamation.

\hypertarget{lock-free-queue-structure}{%
\subsubsection{Lock-free Queue
Structure}\label{lock-free-queue-structure}}

Each queue node contains four critical fields that together provide dual
protection. The \texttt{cycle} field is an immutable temporal
identifier; it is set at enqueue and never changes. The \texttt{next}
field maintains the linked list structure required for FIFO ordering,
with atomic updates ensuring that modifications are visible consistently
across threads and set to \texttt{NULL} on reclaim to prevent stale
traversal. The \texttt{data} field stores the user payload and is only
accessed once the node has been safely claimed. The \texttt{state} field
implements the state-based protection mechanism, transitioning through
\texttt{AVAILABLE} → \texttt{CLAIMED} to guarantee that no node is
reclaimed while still available for dequeue operations.

All linked-list nodes (not payloads) are allocated and recycled from a
type-stable memory pool - nodes reside in a persistent pool, recycled
exclusively as Node objects, and never freed to the OS. This type
stability ensures any pointer to pool memory always references a valid
Node structure with an accessible cycle field, enabling safe cycle-based
protection checks even on recycled addresses.

The \texttt{head} pointer always references a dummy node, simplifying
insertion and deletion. The \texttt{tail} pointer optimizes concurrent
enqueue operations under contention.

The \texttt{scan\_cursor} serves as a dequeue optimization, pointing to
the first likely available node to reduce traversal to \texttt{O(1)} in
the common case. Each successful dequeue advances the cursor atomically
for the benefit of subsequent operations. The
\texttt{protection\_window} parameter bounds memory usage by defining
the temporal window for reclamation safety.

\hypertarget{cycle-management-structure}{%
\subsubsection{Cycle Management
Structure}\label{cycle-management-structure}}

The cycle mechanism enforces temporal safety through invariants rather
than coordination, using two counters to define ordering and reclamation
boundaries.

The global \texttt{cycle} establishes ordering across all enqueue
operations. Each enqueued node is tagged with the current cycle,
creating an immutable temporal identity that enables safe reclamation
decisions.

The \texttt{deque\_cycle} counter tracks the highest cycle value of any
node successfully claimed and processed, representing the current
frontier of dequeue progress. Every dequeue operation unilaterally
announces its progress by updating the global \texttt{deque\_cycle} to
the cycle of the node it has claimed. This announcement requires no
coordination or acknowledgment and always advances monotonically.

Safe reclamation is determined by computing protection window
\texttt{P}. Only nodes outside this window and in state \texttt{CLAIMED}
are reclaimed. This creates a temporal barrier: any node inside
protection window \texttt{P} is guaranteed to be protected. Since
\texttt{deque\_cycle} monotonically tracks the highest cycle accessed by
any dequeue, and the reclaim process maintains a trailing protection
window of size \texttt{W}, this ensures that nodes cannot be freed while
active and cannot be freed if their cycle lies within the protection
window, even if retired.

The 64-bit cycle counter provides a practically infinite sequence space.
At one billion operations per second, wraparound would require more than
584 years, making it irrelevant for practical systems.

\hypertarget{lock-free-enqueue-algorithm}{%
\subsection{Lock-Free Enqueue
Algorithm}\label{lock-free-enqueue-algorithm}}

The enqueue operation implements a modified Michael \& Scott linked-list
insertion algorithm with cycle-based versioning for memory safety.
Multiple producer threads can enqueue concurrently without coordination,
maintaining strict FIFO ordering through atomic linked-list operations.

\begin{algorithm}
\caption{Lock-Free Enqueue}
\begin{algorithmic}[1]
\Function{enqueue}{queue, data}
   \State \textit{// Phase 1: Node allocation and cycle assignment}
   \State $new\_node \gets$ allocate\_node$()$
   \State $new\_node.data \gets data$
   \State $new\_node.next \gets$ NULL
   \State $new\_node.state \gets$ AVAILABLE
   \State $cycle \gets$ INCREMENT$(queue.cycle)$
   \State $new\_node.cycle \gets cycle$
   
   \State \textit{// Phase 2: Lock-free Insertion}
   \State $retry\_count \gets 0$
   \Repeat
       \State $tail \gets$ LOAD$(queue.tail)$
       \State $next \gets$ LOAD$(tail.next)$
       \State \textit{// Pre-check to avoid expensive CAS (OPTIONAL)}
       \If{$next \neq$ NULL}
           \State \textit{// Tail has advanced, retry with fresh state}
           \State $retry\_count \gets retry\_count + 1$
           \If{$retry\_count > 3$}
               \State CPU\_PAUSE$()$
           \EndIf
           \State \textbf{continue}
       \EndIf
       \State \textit{// Attempt to link new node}
       \If{CAS$(tail.next,$ NULL$, new\_node)$}
           \State \textit{// Optional tail advancement}
           \State CAS$(queue.tail, tail, new\_node)$
           \State \textbf{break}
       \EndIf
   \Until{false}
   
   \State \textit{// Phase 3: Conditional Reclamation}
   \If{$(cycle \bmod N) = 0$}
       \State trigger\_reclamation$(queue)$
   \EndIf
   \State \Return SUCCESS
\EndFunction
\end{algorithmic}
\end{algorithm}

\textbf{Phase 1: Node Allocation and Cycle Assignment.} A new node is
allocated from the available pool, and initialized with user data. The
global cycle counter is atomically incremented and assigned to the node,
establishing its unique temporal identity within the 64-bit space. This
ordering enables cycle-based protection, where garbage collection
retains a trailing protection window relative to dequeue progress. If
allocation fails, the algorithm triggers immediate reclamation and
retries, providing automatic memory pressure relief without external
intervention.

\textbf{Phase 2: Lock-Free Insertion.} The insertion phase uses a
streamlined version of the Michael \& Scott algorithm that eliminates
helping mechanisms. The algorithm loads the current \texttt{tail}
pointer and its \texttt{next} field using ACQUIRE ordering to ensure
visibility of concurrent updates. The \texttt{tail-\textgreater{}next}
pointer is expected to be \texttt{null} for normal operation. However,
if the tail's next pointer is non-null, indicating that another thread
has inserted a node, the algorithm retries with fresh state rather than
attempting complex coordination. It uses cpu pause when necessary. When
the tail's next is null, the algorithm attempts to link the new node
using CAS with RELEASE ordering, ensuring all node writes are visible
before publication. After successful insertion, the algorithm advances
the tail pointer to the new node.

\textbf{Phase 3: Conditional Reclamation.} After every N insertions, the
enqueue thread triggers reclamation (cycle \% N == 0), distributing
reclamation work across producers and avoiding dedicated GC threads
while maintaining bounded memory use. The algorithm is agnostic to the
triggering policy - implementations may use deterministic modulo (as
shown), randomized triggers (Bernoulli p = 1/N), or hybrid approaches,
depending on workload fairness requirements. Reclamation is
non-blocking. If another thread is already reclaiming, enqueue proceeds
without reclamation.

\textbf{Performance.} In the common case, enqueue requires 3--5 atomic
operations: one increment for cycle assignment, one or two for loads,
one CAS to link the node, and optionally one CAS to advance the tail.
This matches or improves upon Michael \& Scott's performance while
adding stronger correctness guarantees through cycle-based protection.

\hypertarget{modifications-to-michael-scotts-algorithm}{%
\subsection{Modifications to Michael \& Scott's
Algorithm}\label{modifications-to-michael-scotts-algorithm}}

CMP modifies the original Michael \& Scott (M\&S) algorithm to address
specific inefficiencies, particularly by reducing unnecessary atomic
operations.

\begin{algorithm}
\caption{Original M\&S Helping Mechanism}
\begin{algorithmic}[1]
\Repeat
   \State $tail \gets$ LOAD$(queue.tail)$
   \State $next \gets$ LOAD$(tail.next)$
   \State \textit{// Revalidate tail}
   \If{$tail =$ LOAD$(queue.tail)$}
       \If{$next \neq$ NULL}
           \State \textit{// Help advance tail using 'next' (may be stale)}
           \State CAS$(queue.tail, tail, next)$
           \State \textbf{continue}
       \EndIf
   \EndIf
\Until{false}
\end{algorithmic}
\end{algorithm}

In the original M\&S algorithm, threads attempt to advance the tail
pointer when it is stale (i.e., when next is not null). However, this
helping based on stale data is redundant and often counterproductive.
Under high load, it can increase contention when multiple threads
attempt to act on outdated observations. A more effective approach is to
simply retry with fresh state whenever stale conditions are detected
(Algorithm 1, line 15), ensuring that operations always proceed on
current information. Eliminating ``helping'' reduces both the number of
atomic operations and cache line bouncing in contended scenarios, while
preserving the correctness guarantees of the original algorithm.

Removing this helping mechanism also helps with removing an extra
validation (Algorithm 2 line 5) which was originally meant to avoid
expensive CAS. Since the \texttt{next} non-null value is now eliminated
in earlier condition, the extra atomic validation is no longer required.

These modifications do not compromise any correctness properties of the
original algorithm. The helping mechanism was intended as a performance
optimization, not as a correctness requirement. By removing it, CMP
strengthens correctness by avoiding potential races between stale
observations and helping actions.

It is worth noting that production implementations have taken different
approaches: Java's ConcurrentLinkedQueue employs opportunistic helping,
Boost.Lockfree implements the full helping mechanism as in the original
Michael \& Scott algorithm, while .NET's ConcurrentQueue omits helping
entirely.

\hypertarget{lock-free-dequeue-algorithm}{%
\subsection{Lock-Free Dequeue
Algorithm}\label{lock-free-dequeue-algorithm}}

The dequeue operation follows a coordination-free model where temporal
protection mechanisms are preemptively established, enabling seamless
lock-free execution without requiring explicit safety coordination. It
uses the \texttt{scan\_cursor} optimization to achieve \texttt{O(1)}
common case performance while allowing multiple threads to improve
shared state without explicit coordination.

\begin{algorithm}
\caption{Lock-Free Dequeue}
\begin{algorithmic}[1]
\Function{dequeue}{queue}
    \State $current \gets$ LOAD$(queue.head)$ \Comment{Non-NULL value}
    \State $last\_deque\_cycle \gets 0$
    
    \While{$current \neq$ NULL}
        \State \textit{// Phase 1: Load scan cursor - starting position}
        \State $deque\_cycle \gets$ LOAD$(queue.deque\_cycle)$
        \If{$deque\_cycle \neq last\_deque\_cycle$}
            \State $last\_deque\_cycle \gets deque\_cycle$
            \State $current \gets$ LOAD$(queue.scan\_cursor)$
            \State $last\_cursor \gets current$
            \State $cursor\_cycle \gets last\_cursor.cycle$
        \EndIf
        
        \State \textit{// Phase 2: Atomic node claiming}
        \State $expected \gets$ AVAILABLE
        \If{\textbf CAS$(current.state, expected,$ CLAIMED$)$}
            \State \textbf{break}
        \EndIf
            
	\State $current \gets current.next$
    \EndWhile
    \State if {$current =$ NULL} \textbf{then} \Return NULL
    \State \textit{// Phase 3: Claim Data with CAS}
    \If{$current.state =$ AVAILABLE}
        \State \Return NULL  \Comment{Detect ABA/reassignment}
    \EndIf
    
    \State $data \gets current.data$
    \If{$data =$ NULL \textbf{or not} CAS$(current.data, data,$ NULL$)$}
        \State \Return NULL
    \EndIf
    
    \State $advance\_boundary \gets$ true
    
    \State \textit{// Phase 4: Scan\_cursor Advance \& Cycle Announce}
    \If{$last\_cursor =$ LOAD$(queue.scan\_cursor)$ \textbf{and} 
    \Statex \hspace{2em} $cursor\_cycle = queue.scan\_cursor.cycle$}
        \State $next \gets$ LOAD$(current.next)$
        \State $advance\_boundary \gets$ false
        
        \If{$next =$ NULL \textbf{or} ($next \neq$ NULL \textbf{and}
        \Statex \hspace{2em} CAS$(queue.scan\_cursor, last\_cursor, next))$}
            \State $advance\_boundary \gets$ true
        \EndIf
    \EndIf
    
    \State \textit{// Phase 5: Protection Boundary Update}
    \If{$advance\_boundary$}
        \State $cycle \gets$ LOAD$(queue.deque\_cycle)$
        \While{$cycle < current.cycle$}
        \If{CAS$(queue.deque\_cycle, cycle, current.cycle)$}
            \State \textbf{break}
        \EndIf
        \State $cycle \gets$ LOAD$(queue.deque\_cycle)$
        \EndWhile
    \EndIf
    
    \State \Return $data$
\EndFunction
\end{algorithmic}
\end{algorithm}

\textbf{Phase 1: Scan Cursor Load.} The dequeue loads
\texttt{scan\_cursor}, which points to the first potentially available
node, avoiding re-traversal of already-processed nodes and providing
\texttt{O(1)} performance in most cases without requiring coordination
protocols. The cursor is initialized to the dummy node at setup and is
never set to NULL. During traversal, if \texttt{deque\_cycle} advances
(indicating other threads' progress), the algorithm reloads
\texttt{scan\_cursor} to accelerate convergence under high contention to
reduce traversal.

\textbf{Phase 2: Atomic Node Claiming.} Starting at
\texttt{scan\_cursor}, the thread linearly probes and attempts
\texttt{CAS\ (state,\ AVAILABLE\ →\ CLAIMED)}. This atomic operation
ensures that exactly one thread claims a node; others automatically
continue scanning forward to subsequent nodes. If a CAS fails, another
thread claimed it; continue scanning. On success, the thread has
exclusive access and can read the payload.

\textbf{Phase 3: Atomic Data Claiming.} After claiming, the thread
revalidates that the node state remains \texttt{CLAIMED} (detecting
potential ABA/reassignment scenarios), and extracts data using
CAS(current.data, data, NULL) to atomically claim ownership of the
payload. This CAS prevents duplicate data extraction when multiple
threads might contest for the same data if threads were preempted and
stalled in the \texttt{CLAIMED} state. The atomic nullification ensures
data will be accessed by only one thread, preventing duplicate access
and potential UAF by clearing the data pointer immediately after
extraction.

\textbf{Phase 4: Scan Cursor Advance \& Cycle Announcement.} After
claiming, the thread opportunistically advances \texttt{scan\_cursor} if
and only if both the pointer value and cycle still equal the last loaded
cursor, and the claimed node has a non-null next. This dual condition
mathematically eliminates ABA problems. Since cycles are monotonic and
never wrap around, even if \texttt{scan\_cursor} is reassigned to the
same pointer value, the different cycle will prevent this condition from
being satisfied, providing mathematical certainty against ABA hazards.
This conditional advance prevents redundant atomic operations when
another thread has already advanced the cursor, and if the advancement
CAS fails, correctness is unchanged with only performance impact.

\textbf{Phase 5: Protection Boundary Update.} Upon successful
advancement, it announces temporal progress by writing
\texttt{deque\_cycle\ =\ claimed.cycle}. This maintains the invariant
\texttt{scan\_cursor.cycle} \(\geq\) \texttt{deque\_cycle} and tightens
the protection boundary.

\textbf{Correctness Guarantees.} The Boundary Update (Phase 5) maintains
a trailing protection window that prevents freeing nodes within
protection window of any completed dequeue. The \texttt{scan\_cursor}
optimization maintains correctness even under high concurrency because
all updates move the pointer forward monotonically, and race conditions
only affect performance, never safety. Dual protection ensures nodes
cannot be freed in the state \texttt{AVAILABLE} (state protection) and
cannot be freed while within the temporal window (cycle protection).

\textbf{Performance.} The dequeue requires 4-9 atomic operations in the
common case: one for \texttt{deque\_cycle} loading, one for
\texttt{scan\_cursor} loading, one for node claiming, one for data
claiming, and optionally 3-5 additional operations for
\texttt{scan\_cursor} advancement and \texttt{deque\_cycle} updates. The
\texttt{scan\_cursor} optimization typically reduces scanning to a
single iteration, achieving near-constant time performance regardless of
queue history. Advancement failures are benign---they simply indicate
another thread already optimized the cursor position.

\hypertarget{coordination-free-memory-reclamation}{%
\subsection{Coordination-Free Memory
Reclamation}\label{coordination-free-memory-reclamation}}

\textbf{Safety Predicate.} A node is reclaimed if and only if: \[
(state \neq AVAILABLE) \land (node.cycle < safe\_cycle)
\]

\begin{itemize}
\tightlist
\item
  \texttt{state} \(\neq\) \texttt{AVAILABLE} ensures the node has been
  removed from the abstract queue.
\item
  \texttt{node.cycle\ \textless{}\ safe\_cycle} ensures no active
  consumer thread can access this node or any older cycle.
\end{itemize}

Both conditions are jointly necessary; omitting either risks freeing
memory still reachable by some thread.

The reclamation operates without coordination and is lock-free, allowing
normal queue operations to proceed unimpeded, and uses batched
collection to amortize atomic costs while maintaining predictable
latency. Dual safety conditions ensure nodes are reclaimed only when
both state-based and cycle-based protection mechanisms confirm safety.

Unlike traditional approaches where stalled threads can block garbage
collection indefinitely, CMP allows reclamation to continue past nodes
in \texttt{CLAIMED} state from failed or stalled threads. Once a thread
successfully claims a node, that node becomes a reclamation candidate
after a minimum of \texttt{W} dequeue cycles have passed, enabling
automatic recovery while maintaining all safety guarantees.

\begin{algorithm}
\caption{Coordination-Free Memory Reclamation}
\begin{algorithmic}[1]
\Function{reclaim}{queue}
   \State \textit{// Phase 1: Calculate current protection boundary}
   \State $cycle \gets$ LOAD$(queue.deque\_cycle)$
   \State $window \gets queue.protection.window\_size$
   \State $safe\_cycle \gets \max(0, cycle - window)$
   \State $head \gets$ LOAD$(queue.head)$
   \State $current \gets$ LOAD$(head.next,)$
   
   \While{$current \neq$ NULL}
       \State $original\_next \gets current$
       \State $new\_next \gets current$
       
       \State \textit{// Collect batch of safely reclaimable nodes}
       \While{$current \neq$ NULL}
           \State \textit{// Phase 2: Cycle-based protection check}
           \If{$current.cycle \geq safe\_cycle$}
               \State \textbf{break}
           \EndIf
           \State \textit{// Phase 3: State-based protection check}
           \If{LOAD$(current.state)$ = AVAILABLE}
               \State \textbf{break}
           \EndIf
           
           \State \textit{// Phase 4: Add to reclamation batch}
	   \State $batch.add(current)$
           \State $next \gets$ LOAD$(current.next)$
           \State $new\_next \gets next$
           \State $current \gets next$
       \EndWhile
       
       \State \textit{// Enforce minimum batch size for efficiency}
       \If{$batch.count < MIN\_BATCH\_SIZE$}
           \State \textbf{break}
       \EndIf
       
       \State \textit{// Phase 5: Atomic head pointer advancement}
       \If{CAS$(head.next, original\_next, new\_next)$}
	   \State \textit{// Free and reset nodes (set next = NULL)}
           \State deallocate\_nodes$(batch)$
       \Else
           \State \textit{// Concurrent modification detected}
           \State \textbf{break}
       \EndIf
   \EndWhile
\EndFunction
\end{algorithmic}
\end{algorithm}

\textbf{Phase 1: Protection Boundary Calculation.} Reclamation begins by
calculating the \texttt{safe\_cycle} value that is safe for reclamation
based on the current \texttt{deque\_cycle} value and the configured
protection window size \texttt{W}. \[
safe\_cycle = deque\_cycle - W
\] Nodes with cycles below this \texttt{safe\_cycle} threshold are
candidates for reclamation, while nodes inside the protection window are
preserved to prevent ABA hazards. Only nodes in \texttt{CLAIMED} state
are considered; \texttt{AVAILABLE} nodes are absolutely protected
regardless of cycle value.

\textbf{Phase 2: Cycle-Based Safety Boundary.} Starting from
\texttt{head.next}, the algorithm first checks whether a node's cycle is
below the threshold. Since the \texttt{cycle} field is immutable, this
check is a fast, non-atomic read. Only nodes passing this test proceed
to state verification. Dequeue operations maintain the invariant
\texttt{scan\_cursor.cycle} \(\geq\) \texttt{deque\_cycle}, and the
\texttt{tail} always holds the latest cycle value, so active queue
boundaries remain protected without explicit pointer checks.

\textbf{Phase 3: State-Based Safety.} The algorithm confirms that the
node is not in \texttt{AVAILABLE} state. Nodes in \texttt{AVAILABLE}
state are never reclaimed, regardless of their cycle values, as they
represent nodes waiting to be dequeued. Reclamation halts at the first
node that is in \texttt{AVAILABLE} state to maintain FIFO order and
prevent premature reclamation.

\textbf{Phase 4: Batch Collection.} Eligible nodes are collected in a
batch rather than reclaimed one by one. Batching avoids repeated head
updates, reduces atomic operations, and improves cache efficiency. The
algorithm records the original head and the new head after all
reclaimable nodes, performing a single atomic CAS to advance across the
batch.

\textbf{Phase 5: Atomic Head Update and Memory Management.} When the
batch reaches the minimum threshold, reclamation attempts a CAS update
of the head pointer. If another thread interferes, the attempt is
abandoned to avoid consistency issues. Only after a successful CAS are
the collected nodes processed for deallocation. To ensure any dequeue
thread with a stale pointer safely terminate traversal, the
\texttt{next} and \texttt{data} pointers set to NULL before being
returning free node to the memory pool.

\textbf{Mathematical Safety Guarantees.} Temporal progression is
enforced by monotonic updates of \texttt{deque\_cycle}. Reclamation lags
by at least \texttt{W} cycles, ensuring that any active or stalled
thread remains protected against reclamation. This eliminates UAF and
ABA hazards without coordination, provided the protection window
\texttt{W} exceeds the maximum possible delay in dequeue progress.

\hypertarget{correctness-and-progress-guarantees-2}{%
\subsection[Correctness and Progress Guarantees
]{\texorpdfstring{Correctness and Progress Guarantees \footnote{Complete
  formal proofs of linearizability, FIFO ordering, memory safety,
  lock-free progress, and bounded reclamation properties available upon
  request.}}{Correctness and Progress Guarantees }}\label{correctness-and-progress-guarantees-2}}

CMP achieves strict FIFO semantics, lock-free progress, unbounded
capacity, and bounded reclamation, with common-case \texttt{O(1)}
latency and stable performance under contention. The design simplifies
the enqueue path by eliminating ``helping'' and redundant validations
from M\&S, and ensures automatic recovery from failed or stalled threads
through its bounded temporal protection window.

\textbf{Linearizability.} Enqueue operations linearize at the successful
\texttt{CAS(tail.next,\ null,\ new\_node)} that links a node, while
dequeue operations linearize at the successful
\texttt{CAS\ (current.state,\ AVAILABLE,\ CLAIMED)} that claims
ownership. Empty dequeues linearize when the scan cursor reaches null.
These points yield a total order consistent with real-time precedence
and sequential queue semantics.

\textbf{FIFO Ordering.} Strict FIFO ordering is maintained through three
key invariants: (1) \textbf{Append-only linking:} Enqueues link new
nodes strictly at the physical tail via
\texttt{CAS\ (tail.next,\ null,\ new\_node)}, establishing chronological
order. After linking, list order is immutable. (2) \textbf{Cursor
minimality:} The \texttt{scan\_cursor} never advances past an
\texttt{AVAILABLE} node---every node strictly before
\texttt{scan\_cursor} is in \texttt{CLAIMED} state. (3)
\textbf{Earliest-claim property:} Dequeues linearly probe from
\texttt{scan\_cursor}; a node can be claimed
\texttt{(AVAILABLE\ →\ CLAIMED)} only after all predecessors are
non-AVAILABLE. Together, (1)--(3) ensure the first successful claim
always targets the earliest available node in list order, guaranteeing
removals respect enqueue order regardless of thread scheduling.

\textbf{Lock-Free Progress.} All operations complete in bounded time
with system-wide progress guarantees. Enqueues use simplified M\&S
insertion where each failed CAS indicates another thread's successful
progress. Dequeues scan forward from optimized positions, with each
failed claim CAS representing another thread's advancement. The
\texttt{scan\_cursor} optimization achieves \texttt{O(1)} amortized
dequeue performance by tracking the earliest viable position.

\textbf{Bounded Reclamation.} Nodes are reclaimed within at most
\texttt{W} dequeue cycles plus GC delay cycles after reaching
\texttt{CLAIMED} state, independent of thread failures, scheduling
stalls, or queue size. This provides predictable memory usage, stable
latency, and resilience under crashes.

\textbf{Memory Safety Guarantees.} The state-based protection prevents
reclamation of nodes in the \texttt{AVAILABLE} state, while the
cycle-based protection retains processed nodes within a bounded temporal
window \texttt{P}. Together, these mechanisms ensure absolute safety
against UAF, and safety against ABA within the configured window size
\texttt{W}, without relying on complex retries, validation, or
coordination overhead, achieving both safety and performance through
algorithmic invariants rather than runtime detection.

These properties are established through rigorous mathematical analysis
including invariant preservation, progress arguments, and safety proofs.
The formal treatment demonstrates that CMP achieves the challenging
combination of strict FIFO semantics, high performance,
coordination-free operation, and automatic fault recovery without
sacrificing correctness guarantees.

\hypertarget{performance-evaluation}{%
\section{Performance Evaluation}\label{performance-evaluation}}

We evaluate CMP against production-ready lock-free queue implementations
deployed at scale to assess performance and scalability under real-world
conditions. CMP was originally designed to meet production requirements
in mesibo (real-time communication platform) and PatANN
(high-performance vector search engine), both requiring unbounded,
coordination-free queues to sustain high-throughput, low-latency
workloads.

Numerous academic prototypes (LCRQ \citep{morrison2013fast}, FAA-based
queues \citep{yang2016wait}, and others) were excluded, as they either
rely on non-portable primitives (e.g., DCAS or architecture-specific
atomics unavailable on ARM and in managed languages) or remain research
prototypes with no production adoption or large-scale deployment, making
them unsuitable for representing real-world industry scenarios.

\textbf{CMP Queue:} The CMP implementation with strict FIFO ordering and
unbounded queue.

\textbf{Moodycamel (MC) Concurrent Queue \citep{moodycamel_queue} }:
Industry-standard lock-free queue implementation employing a relaxed
ordering model that sacrifices strict FIFO semantics to maximize
throughput, with optimized fast paths for single-producer,
single-consumer settings.

\textbf{Boost Lockfree Queue \citep{boost_lockfree}:} Based on the
Michael \& Scott (M\&S) algorithm, using hazard pointers for memory
safety and CAS for synchronization. It provides strict FIFO and serves
as a baseline. CMP also builds on M\&S but replaces hazard pointers with
cyclic memory protection, showing that modern reclamation schemes can
improve both performance and simplicity.

All experiments were conducted on a controlled testbed with round-robin
sequencing of implementations to eliminate bias from CPU thermal
throttling and dynamic frequency scaling. To reduce noise from OS
preemption, interrupts, and memory traffic, we applied 3-sigma filtering
uniformly across all implementations (CMP, Moodycamel, Boost): samples
beyond \(\mu\) \(\pm\) 3\(\sigma\) were discarded, removing
\textasciitilde{}0.3\% of anomalies and yielding stable, reproducible
results. This standard practice \citep{georges2007statistically} ensures
P99 values reflect queue performance under stable conditions rather than
random OS or hardware artifacts. We report two regimes: baseline,
measuring raw queue throughput; and synthetic load, where threads
perform additional computation between operations to emulate realistic
workloads.

\hypertarget{performance-results}{%
\subsection{Performance Results}\label{performance-results}}

\hypertarget{throughput-performance}{%
\paragraph{Throughput Performance}\label{throughput-performance}}

Throughput under different producer-consumer configurations is shown in
Table 1.

\begin{figure}[htbp]
    \centering
    \vspace{-\baselineskip}  % Remove one line of space at top
    \caption{Throughput comparison across thread configurations. CMP maintains superior performance across all configurations, with particularly dramatic advantages at high contention (64P64C shows 425\% improvement over alternatives).}
    \label{fig:cmp_performance}
    \includegraphics[width=\columnwidth]{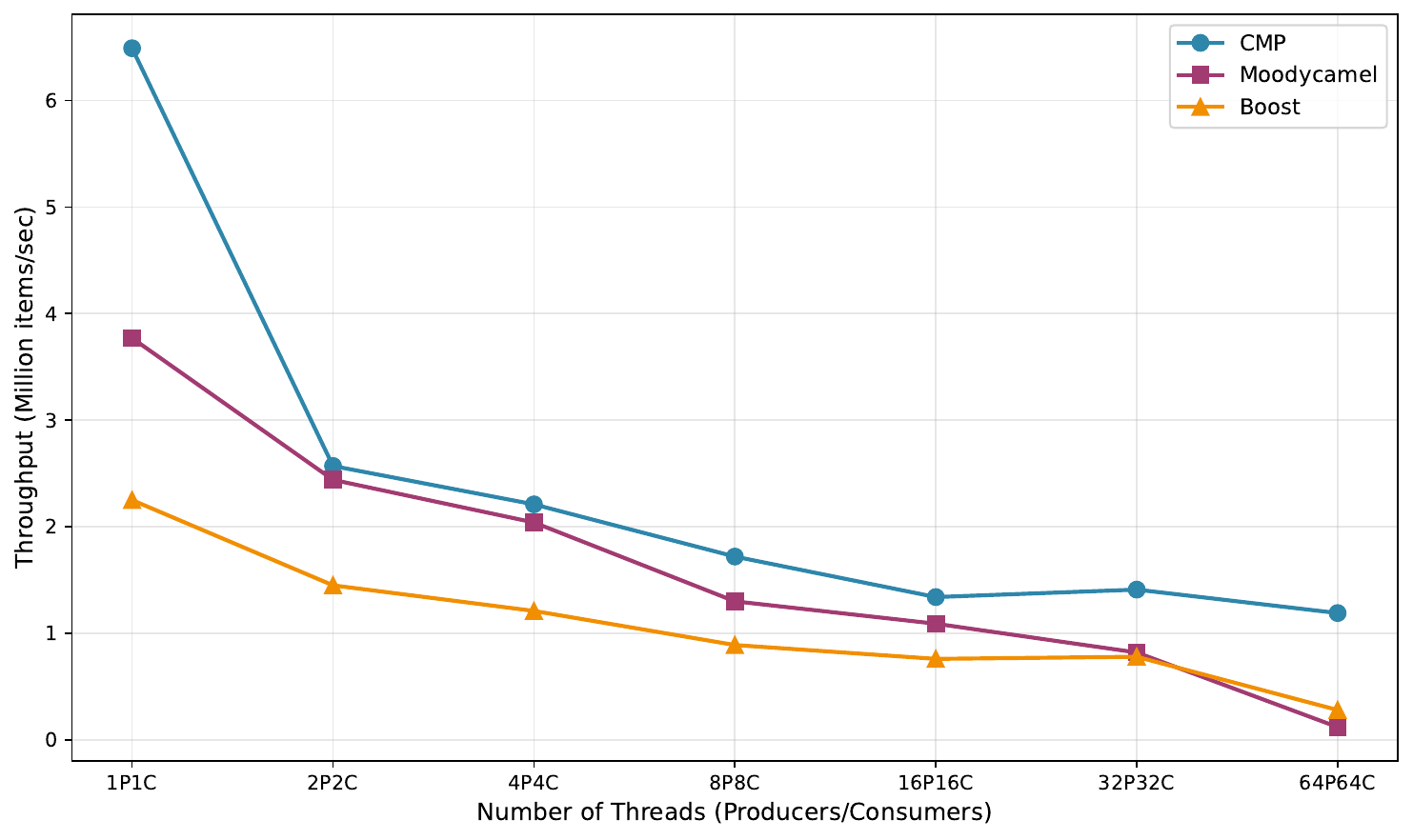}
\end{figure}

In the single-producer/single-consumer case (1P1C), CMP sustains 6.49M
items/second, 72\% higher than Moodycamel and 188\% higher than Boost,
reflecting CMP's reduced synchronization overhead when contention is
minimal. At the largest configuration (64P64C), CMP reaches 1.19M
items/second, 325\% higher than Boost and 892\% higher than Moodycamel,
indicating exceptional scaling under extreme contention. Notably, at
this extreme contention level, Boost outperforms Moodycamel by 2.3x in
throughput, suggesting that Moodycamel's coordinated design suffers
badly under maximum stress---a pattern that underscores the importance
of CMP's coordination-free approach.

For intermediate contention (2P2C---16P16C), CMP achieves 5---23\%
higher throughput than Moodycamel while preserving strict FIFO ordering.
This advantage demonstrates that CMP's coordination-free design not only
maintains ordering but actually improves performance under moderate
contention.

Across all configurations, CMP outperforms Boost by 77---325\%,
demonstrating efficient lock-free design while retaining correctness
properties.

\hypertarget{latency-analysis}{%
\paragraph{Latency Analysis}\label{latency-analysis}}

The following analysis examines operation costs under varying contention
levels, reporting latency in nanoseconds for no contention (1P1C),
balanced contention (4P4C), high contention (32P32C), and extreme
contention (64P64C).

\begin{table}[htbp]
	\caption{Latency with no contention. CMP achieves 40\% lower enqueue and 50\% lower dequeue latencies than Moodycamel (MC).}
\label{tab:latency_performance}
\centering
\begin{tabular}{lcccc}
\toprule
\textbf{Impl} & \textbf{Avg Enq} & \textbf{P99 Enq} & \textbf{Avg Deq} & \textbf{P99 Deq} \\
\midrule
CMP    & \textbf{63.9} & \textbf{111} & \textbf{70.6} & \textbf{74} \\
Moodycamel & 106.0         & 146          & 141.9         & 172          \\
Boost      & 258.5         & 314          & 164.0         & 193          \\
\bottomrule
\end{tabular}
\end{table}

In the single-producer, single-consumer case, CMP achieves the lowest
enqueue and dequeue latencies. Compared to Moodycamel, enqueue
operations are 40\% faster on average, and dequeue operations are 50\%
faster. These results indicate that CMP's coordination-free enqueue and
dequeue mechanisms incur low overhead and latency in the baseline
configuration.

\begin{table}[htbp]
\caption{Under balanced contention, CMP shows higher enqueue latency but 38\% lower dequeue latency than MC.}
\label{tab:contention_latency}
\centering
\begin{tabular}{lcccc}
\toprule
\textbf{Impl} & \textbf{Avg Enq} & \textbf{P99 Enq} & \textbf{Avg Deq} & \textbf{P99 Deq} \\
\midrule
CMP        & 217.1         & 314          & \textbf{87.5} & \textbf{116} \\
Moodycamel & \textbf{144.3} & \textbf{213} & 170.4          & 211          \\
Boost      & 401.8         & 489          & 233.0          & 307          \\
\bottomrule
\end{tabular}
\end{table}

With balanced producer---consumer load, CMP shows higher enqueue latency
(50\% higher than Moodycamel) due to strict FIFO enforcement, but
achieves 49\% lower dequeue latency. This trade-off results in overall
performance advantages: despite slower enqueues, faster dequeues allow
CMP to maintain competitive throughput while providing strict ordering
guarantees.

\begin{table}[htbp]
\caption{Under high contention, CMP reduces enqueue latency by 10\% and dequeue latency by 70\% relative to MC.}
\label{tab:high_contention_latency}
\centering
\begin{tabular}{lcccc}
\toprule
\textbf{Impl} & \textbf{Avg Enq} & \textbf{P99 Enq} & \textbf{Avg Deq} & \textbf{P99 Deq} \\
\midrule
CMP        & \textbf{359.3} & \textbf{350} & \textbf{66.4} & \textbf{70} \\
Moodycamel & 398.7          & 387          & 223.1          & 289          \\
Boost      & 445.2          & 542          & 267.8          & 354          \\
\bottomrule
\end{tabular}
\end{table}

Under high contention (32P32C and 64P64C), CMP consistently reduces both
enqueue and dequeue latencies compared to Moodycamel. At 32P32C, CMP
achieves 10\% lower enqueue latency and 70\% lower dequeue latency,
while at 64P64C it maintains 14\% lower enqueue and 30\% lower dequeue
latencies. CMP also demonstrates superior P99 characteristics across
both configurations, underscoring its robustness under maximum stress
conditions.

\hypertarget{synthetic-workload-resilience-analysis}{%
\paragraph{Synthetic Workload Resilience
Analysis}\label{synthetic-workload-resilience-analysis}}

To assess performance under realistic workloads, we applied a synthetic
mixed workload by interleaving queue operations with additional
computation, inducing memory pressure, cache contention, and scheduling
interference. Retention is reported as the fraction of baseline
throughput sustained under this load.

\begin{figure}[htbp]
    \centering
    \vspace{-\baselineskip}  % Remove one line of space at top
    \caption{Performance retention under synthetic load. CMP maintains 75-92\% throughput across all configurations, demonstrating superior resilience under contention.}
    \label{fig:phantom_load}
    \includegraphics[width=\columnwidth]{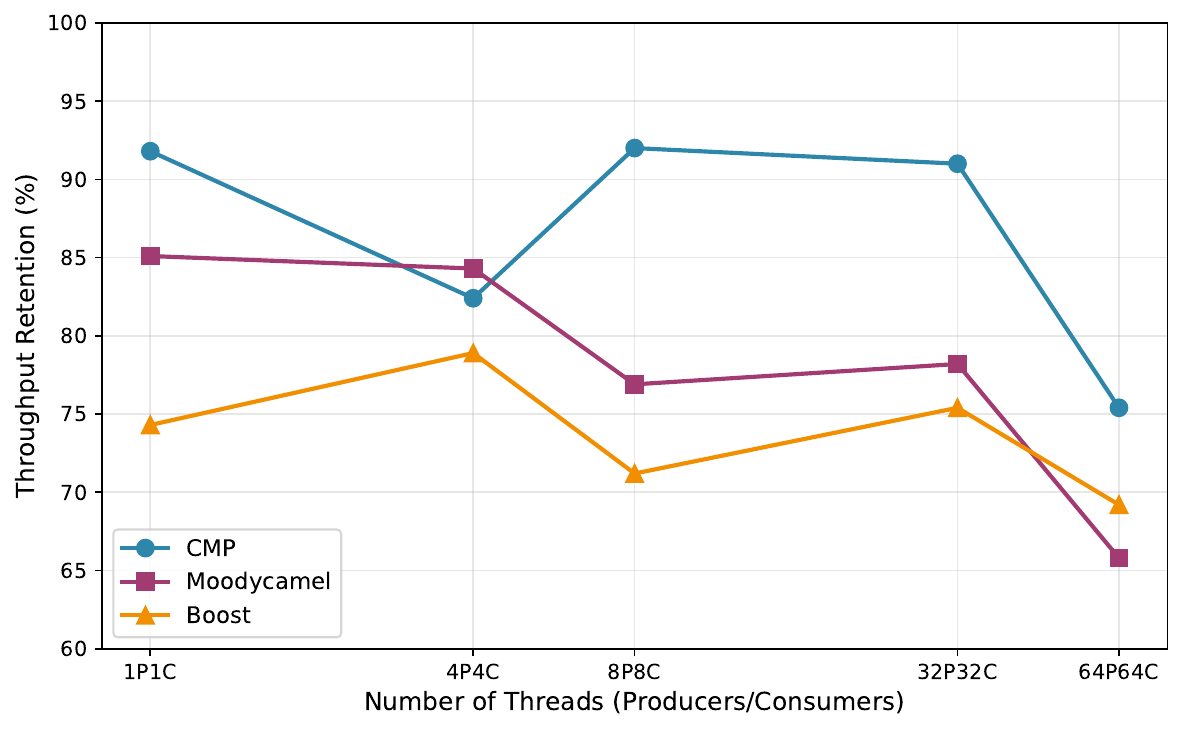}
\end{figure}

CMP maintains 92\% of baseline performance at 8P8C, compared to
Moodycamel's 76.9\%, a +15.1 percentage-point advantage. At 1P1C, CMP
sustains 91.8\% retention with a +6.7 pp advantage.

The strong retention across configurations indicates that CMP's
coordination-free design is less affected by cache pressure and memory
subsystem interference. By eliminating synchronization between producers
and consumers, CMP reduces waiting and scheduling overhead. In contrast,
Boost shows the weakest resilience (69--78\% retention), reflecting the
sensitivity of coordination-based schemes to memory pressure.

\textbf{Performance Summary.} CMP's coordination-free enqueue and
dequeue mechanisms deliver high throughput while enforcing strict FIFO
ordering. In the baseline case (1P1C), CMP achieves 6.49M items/sec,
which is 72\% higher than Moodycamel and 188\% higher than Boost,
showing that FIFO enforcement incurs negligible overhead without
contention. At maximum concurrency (64P64C), CMP sustains 1.19M
items/sec, representing a 325\% improvement over Boost and an 892\%
improvement over Moodycamel, demonstrating scalability under extreme
load.

Across balanced producer--consumer scenarios, CMP remains competitive or
superior across all metrics while preserving deterministic FIFO
guarantees. Relative to the academic baseline, CMP outperforms Boost by
77--325\% across configurations, with the largest advantages at high
concurrency levels. Under synthetic load, CMP retains up to 92\% of
baseline throughput, including a +15.1 percentage-point advantage at
8P8C over Moodycamel, underscoring resilience to cache pressure, memory
interference, and scheduling effects.

At 64P64C, Moodycamel's throughput falls below Boost, illustrating that
complex optimization strategies can exhibit scalability limitations
under extreme contention. These results reinforce the benefits of CMP's
coordination-free approach, which combines scalability with strict
correctness guarantees.

\hypertarget{conclusion-and-future-work}{%
\section{Conclusion and Future Work}\label{conclusion-and-future-work}}

CMP demonstrates that strict FIFO ordering, high throughput, and memory
safety can coexist in lock-free queue design. CMP eliminates UAF and ABA
by construction rather than via complex validation protocols, while
maintaining competitive performance across contention regimes. While CMP
already achieves scalability, a segmented variation---similar to
Moodycamel's---could further increase scalability under extreme
contention, while preserving CMP's correctness guarantees and automatic
recovery properties.

%% Bibliography
\bibliography{bibfile}

\end{document}